\documentclass[12pt,twoside,a4paper]{enc}
\usepackage{epsf,exscale,times}
\usepackage[english]{babel}
\usepackage{graphicx}
\usepackage{amsmath}
\usepackage{fancyhdr}
\usepackage[bookmarks=false]{hyperref}

\newcommand {\vsp}   {\vspace*}

\def\title#1{\vsp{-16mm}\begin{center}\Large\bf{#1}\end{center}\vsp{0mm}}
\def\author#1{{\begin{center}\textbf{#1}\end{center}\vspace{-1mm}}}
\def\address#1{\vsp{-3mm}\begin{center}\baselineskip12pt\normalsize{#1}\end{center}\vsp{-1mm}}

\def\abstract#1{{\vspace{-5mm}
    \begin{center}
      \begin{minipage}{0.85\textwidth}
        \noindent\bf \textit{Abstract:}
        \small\rm\emph{#1}
				\vsp{-0.5em}
      \end{minipage}
    \end{center}
}}

\def\authorsheadline=#1{\global\def\@authorsheadline{#1}}
\global\def\@authorsheadline{}

\def\TeX{T\kern-.1667em\lower.5ex\hbox{E}\kern-.125emX}

\def\LaTeXG{{\rm L\kern-.36em\raise.3ex\hbox{\sc a}\kern-.15emT\kern-.1667em\lower.7ex\hbox{E}\kern-.125emX}}
\def\LaTeXK{{\it L\kern-.24em\raise.4ex\hbox{\scriptsize \it A}\kern-.20emT\kern-.1667em\lower.5ex\hbox{E}\kern-.125emX}}

\begin{document}

\fancypagestyle{firststyle}
{
   \fancyhf{}
   \lfoot{ \footnotesize{The European Navigation Conference ENC 2020, November 22-25, 2020, Dresden, Germany\break     978-3-944976-28-0 \copyright  2020 DGON} }
   \rfoot{ \footnotesize {\thepage} }
}

\thispagestyle{firststyle}
\fancyhf{}
\renewcommand{\headrulewidth}{0pt}
\renewcommand{\footrulewidth}{1pt}
\renewcommand{\footskip}{50pt}

\pagestyle{fancy}
\fancyfoot[RO,LE]{ \footnotesize {\thepage} }

\title{Simulation-based Analysis of Multipath Delay Distributions in Urban Canyons}


\author{
Simon Ollander$^{*\dagger} $, Friedrich-Wilhelm Bode$^{\dagger}$, Marcus Baum$^{*}$}
\address{
$^{*}$Institute of Computer Science, University of Goettingen\\ 37077 Goettingen, Germany \\
simon.ollander@stud.uni-goettingen.de\\marcus.baum@cs.uni-goettingen.de\\[2mm]
$^{\dagger}$Robert Bosch Car Multimedia GmbH\\ 31139 Hildesheim, Germany \\
simon.ollander@de.bosch.com \\ fritz.bode@de.bosch.com
}

\abstract{
Global navigation satellite systems provide accurate positioning nearly worldwide. However, in the urban canyons of dense cities, buildings block and reflect the signals, causing multipath errors. To mitigate multipath errors, knowledge of the distribution of the reflection delays is important. Measurements of this distribution have been done in several dense cities, but it is unknown how the delay distribution depends on the depth of the urban canyon. To fill this gap, we simulated reflection scenarios in 12 different environments: from suburban to deep urban canyon. Subsequently, we analyzed the resulting delay distributions. This paper presents these distributions, and a method to estimate them using the number of received satellites. According to our simulation, the multipath delays follow gamma distributions, whose shape parameters decrease when the urban canyon depth increases. A quadratic model can estimate the shape parameter using the number of received satellites. Consequently, depending on the number of received satellites,  the distribution of the reflection delays can be estimated. This information can be combined with prior knowledge from other methods for improved multipath delay estimation. In the future, for more realistic results, the effects on signals that are reflected multiple times and environments other than urban canyons should be simulated.
}


\section{Introduction} 
\begin{figure}[t]
	\centering
	\includegraphics[width=0.45\textwidth]{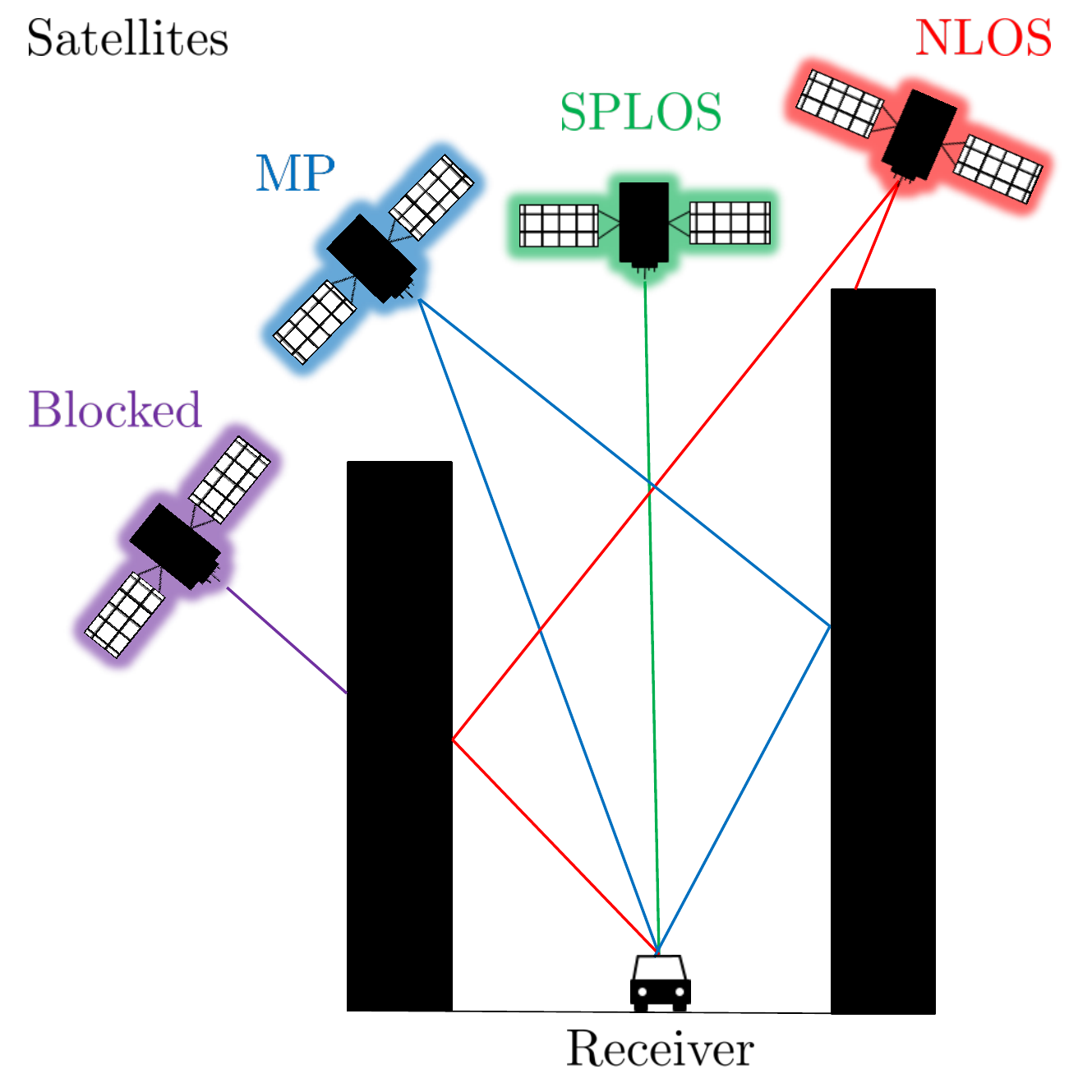}
	\caption{Urban canyons pose three different problems to satellite signals: they can completely block them, allow only a reflection to be received (non-line-of-sight, NLOS), allow the direct signal and a reflection to be received (multipath, MP), or only allow the direct signal (single path line-of-sight, SPLOS). The figure is adapted from~\cite{Olla18}.}
	\label{fig:mpfig}
\end{figure}

Global navigation satellite systems (GNSS) allow their users to estimate their position with high precision, almost world-wide in outdoor areas. Their applications include navigation for vehicles and pedestrians, and the Internet of Things. The main GNSS positioning error is multipath (MP) and non-line-of-sight (NLOS), caused by buildings reflecting and blocking the signals, as depicted in Fig.~\ref{fig:mpfig}. MP and NLOS errors falsify the pseudorange measurements, which are critical to solve the navigation equations and determine the position of the receiver. Numerous methods have been proposed to minimize multipath errors~\cite{Bres16,Zhu18}. One category of these methods is based on detection~\cite{Olla19}, with the purpose of excluding erroneous pseudorange measurements from the positioning algorithm. Excluding satellites is only feasible if at least four error-free (single path line-of-sight, SPLOS) satellites are available. Another category is mitigation~\cite{Olla20_2}, which corrects erroneous pseudorange measurements. To do this, the reflection delay or the pseudorange error has be estimated, which is a complex task. However, the geometrical configuration of the reflecting buildings makes some reflection delay values more probable than others, which can be summarized using a probability density function (PDF). These PDFs can be used as a support for multipath mitigation methods, to be able to assess the environment-based probability of an estimated reflection delay. 

Related work on multipath delay distributions includes two categories: measurements and simulations.

The first method, to measure multipath distributions, requires one to move a receiver capable of measuring reflections around a city~\cite{Stei03, Xie14, Rehm12, Medi18}. An advantage is that the results will be realistic and representative of the used environment. Disadvantages include problems to generalize for new environments, and that the hardware will limit the precision of the reflection measurements. Based on measurements in Shanghai, the delays follow a gamma distribution, where the shape parameter represents the number of times each signal is reflected on average~\cite{Wang18}.

The second method, simulating multipath distributions, requires 3D-modeling of the environment and ray tracing to compute the reflections~\cite{Stei04, Lehn05, Stei09, Stei19, Esbr04}. It has the advantage of allowing free variation of the environment, and it is fairly simple as long as the signals only reflect once. However, if the signals are reflected multiple times (multi-reflection), the simulation becomes computationally heavy.







In this paper, we present the results of a ray tracing simulation, that has been repeated for 12 environments with different mean building heights. As in related work, the reflections follow a gamma distribution, but in addition we show that the shape parameter decreases with increasing building height. Furthermore, a quadratic model is proposed to estimate the shape parameter using the number of available satellites, which is feasible for real-time applications.

\section{Simulations}\label{sec:sim}
Each simulation consists of four steps: urban canyon generation, definition of antenna and satellite trajectories, and ray tracing. All simulations were sampled once per second.

\subsection{Urban Canyon Generation}\label{sec:gen}
\begin{figure}[t]
	\centering
	\includegraphics[width=0.8\textwidth]{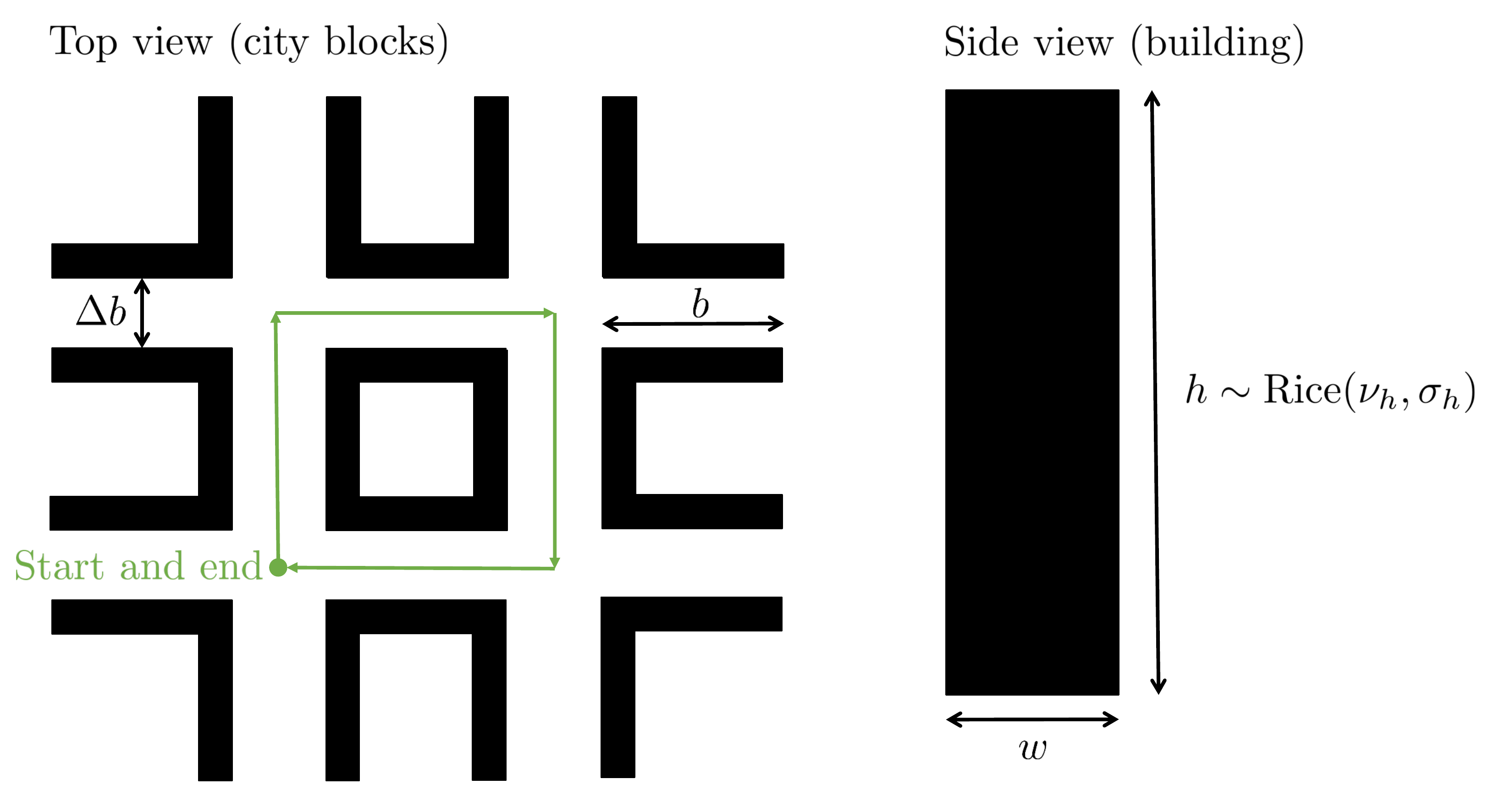}
	\caption{The urban canyon's dimensions are deterministic (left), while the width and height of each building follow a Rice  distribution (right).}
	\label{fig:geo}
\end{figure}

\begin{figure}[t]
	\centering 
	\includegraphics[trim=0 0 0 0,clip,width=0.75\textwidth]{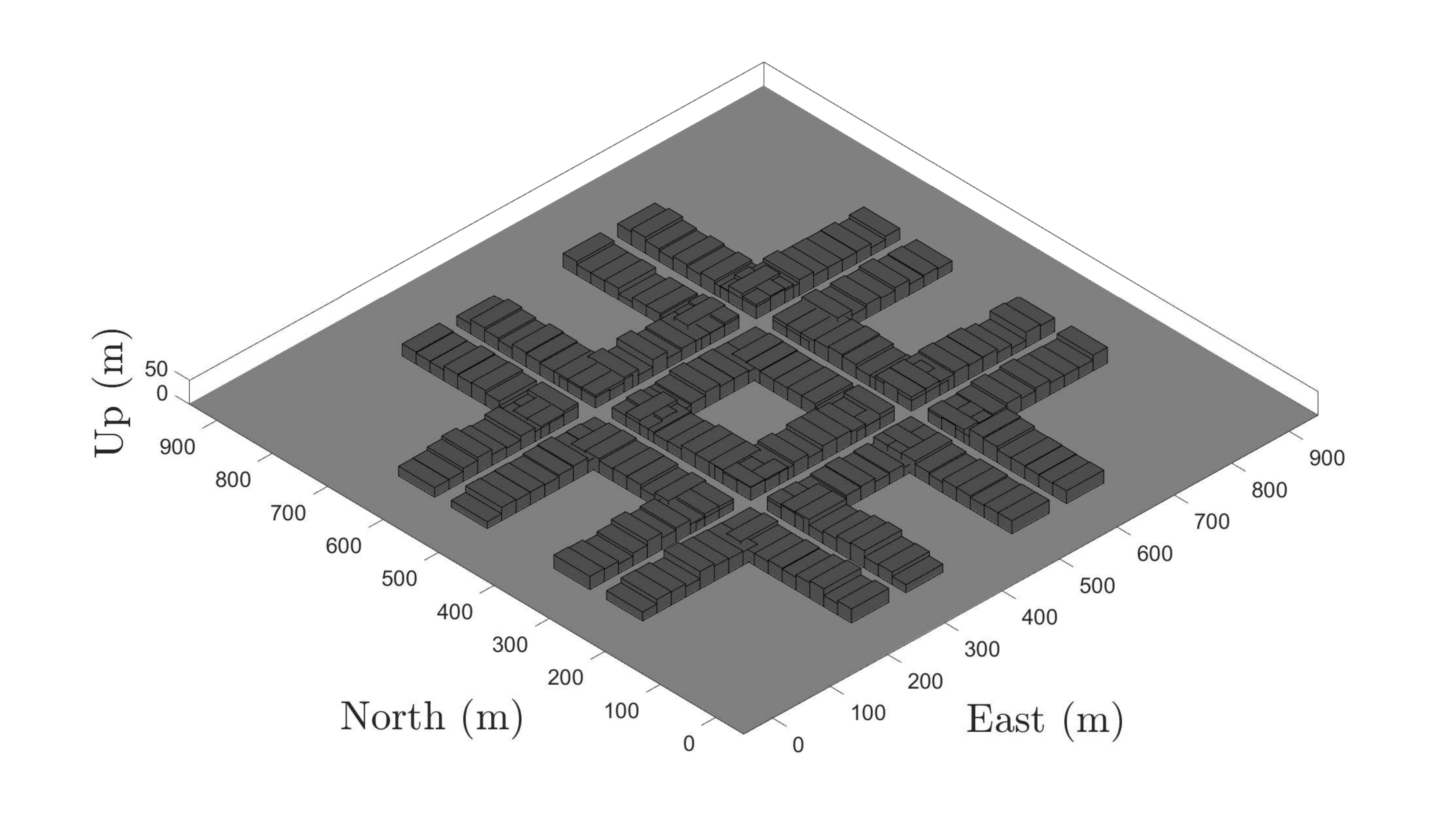}
	\caption{An example of a randomly generated geometry with $\nu_h = 25 \, \text{m}$.}
	\label{fig:vissigmah20}
\end{figure}

\begin{figure}[t]
	\centering
	\includegraphics[trim=100 0 123 0,clip,width=0.85\textwidth]{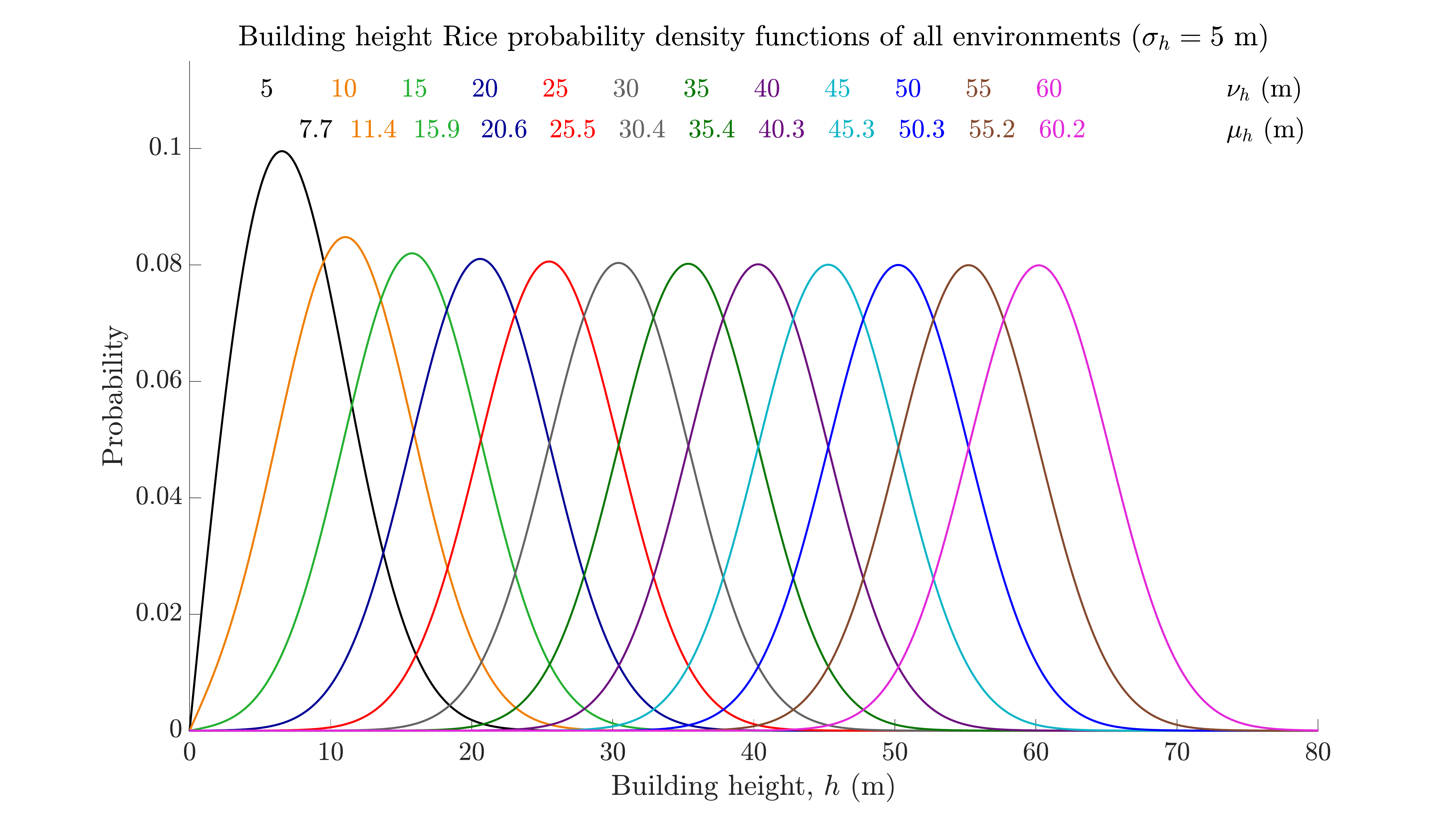} 
	\caption{We simulated 12 environments with different mean building heights $\mu_h$. The mean building height is slightly higher than the $\nu_h$ parameter, since the Rice distribution only supports positive values.}
	\label{fig:nuvalues}
\end{figure}

As a first step, 12 urban environments were generated, each consisting of nine city blocks. The city blocks are squares with side $b = 250 \, \text{m}$, separated by the road width $\Delta b = 30 \, \text{m}$, see Fig.~\ref{fig:geo}. Each building is defined by its width $w = 25 \, \text{m}$ and its height $h$, which follows a Rice distribution~\cite{Rice45}
\begin{align}
h \sim \text{Rice}(\nu_h,\,\sigma_h),
\label{eq:geo}
\end{align}
with non-centrality parameter $\nu_h$ and scale parameter $\sigma_h$. The Rice distribution was chosen because it is strictly positive and because its two parameters allow an adjustment of a mean and a standard deviation similarly to the Gaussian distribution.
In Fig.~\ref{fig:vissigmah20}, a generated geometry with $\nu_h = 25 \, \text{m}$ can be seen.
For the environments, the value of $\nu_h$ varied from $5 \, \text{m}$ to $60 \, \text{m}$ in steps of $ 5 \, \text{m}$, see Fig.~\ref{fig:nuvalues} (all with  $\sigma_h = 5 \, \text{m}$). Since the Rice distribution only supports positive values, its mean 
\begin{align}
\mu_h = \sigma_h\sqrt{\frac{\pi}{2}} L_{1/2}\bigg(\frac{-\nu_h^2}{2\sigma_h^2}\bigg)
\label{eq:muh}
\end{align}
is slightly higher than $\nu_h$, especially for lower values of $\nu_h$. Here, $L_{1/2}$ is the Laguerre polynomial of degree $1/2$. Hereafter, the mean building height $\mu_h$ will be used to represent the mean urban canyon depth (which is equivalent to mean building height).

\subsection{Vehicle and antenna}
The vehicle was modeled as a cuboid with identical sides of $2 \, \text{m}$, and a height of $1.5 \, \text{m}$. The antenna was on its roof, with a distance over the roof of $\delta = 1 \, \text{cm}$. The vehicle moved with the velocity $\bar{v}_a$ and the speed $\| v_a \| = 5 \, \text{m/s}$. At the equator, the satellites reach a higher elevation than at extreme southern or northern latitudes, which is why we used a position near New York City: $\lambda = 40 ^\circ$, $\phi = -70 ^\circ$. At this location, $7.88$ satellites are above $15^{\circ}$ elevation on average, this is what the receiver would receive if the buildings were not there. The vehicle drove in the middle of the road, and performed a full turn around the central city block, see Fig.~\ref{fig:geo}. One simulation represents  $224 \, \text{s}$ and a distance of  $1.12 \, \text{km}$.

\subsection{Satellite orbits}
For each point in time, satellite positions of all 31 GPS satellites were generated. The scenario was repeated six times, each repetition starting two hours later than the previous repetition (since the satellite positions repeat themselves after ca. 12 hours). This fast-forwarding was done to average out different satellite constellations. The elevation of a satellite is $\theta$ and its azimuth angle is $\beta$. 

\subsection{Ray tracing}
\begin{figure}[t]
	\centering
	\includegraphics[trim=0 10 0 5,clip,width=0.35\textwidth]{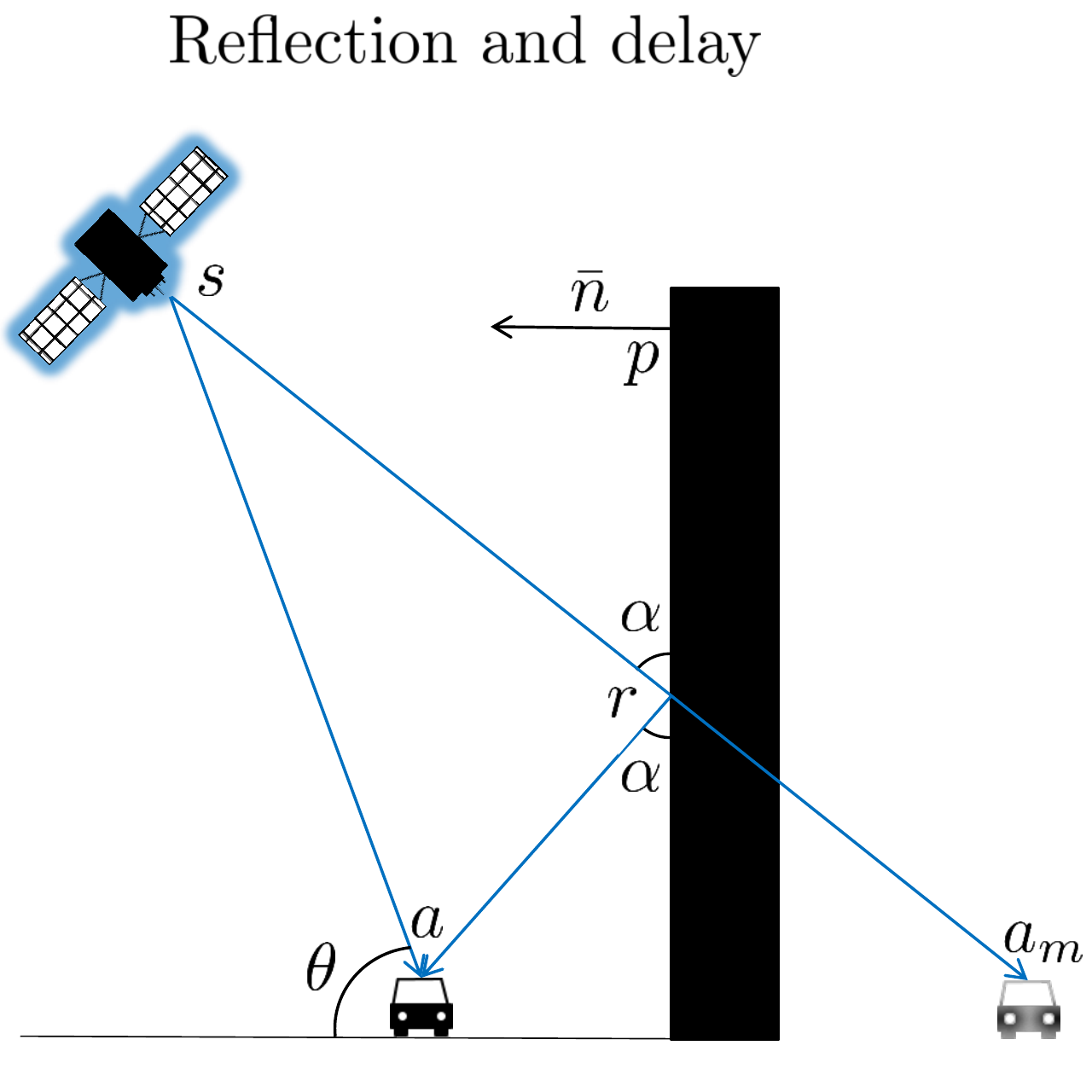}
	\caption{The reflection point $r$ was computed using mirroring.}
	\label{fig:refl}
\end{figure}

For each point in time, given a satellite position $s$ and the position of the vehicle antenna $a$, and given a plane (defined with a point of the plane $p$ and its normal $\bar{n}$), the reflections were computed. First, the mirror point $a_m$ of the antenna position was computed, see Fig.~\ref{fig:refl}:  
\begin{align}
a_m =  a + 2 \bar{n} \bigg(    \frac{( p - a) \cdot \bar{n}}{\bar{n} \cdot \bar{n}}  \bigg).
\label{eq:mirror}
\end{align}  
Second, the reflection point in the plane was computed using projection:
\begin{align}
r = s + \frac{(p-s) \cdot \bar{n}}{(a_m - s) \cdot \bar{n}} (a_m - s).
\label{eq:reflection}
\end{align}  
Third, a check was made whether $r$ was within the limits of the plane, to decide if it really was a reflection point. This was repeated for all the planes that were generated. Finally, the reflection delay $d$ was computed as the difference between the direct path and the path via the reflection point
\begin{align}
d = \|r-s\| + \|a-r\| - \|a-s\|.
\label{eq:delay_def}
\end{align}

\section{Results and Discussion}
This section presents and discusses the results of the simulation. First, the reception modes are presented, followed by the delay distributions, and a model to estimate the delay distributions. The simulations identified a large number of small ($d < \delta = 1 \, \text{cm}$) car roof reflections. These values are too small to be relevant for reflection delay estimation, thus they were excluded  from the subsequent analysis. 

\subsection{Reception modes}\label{sec:recmode}
\begin{figure}[t]
	\centering 
	\includegraphics[trim=70 0 125 0,clip,width=0.8\textwidth]{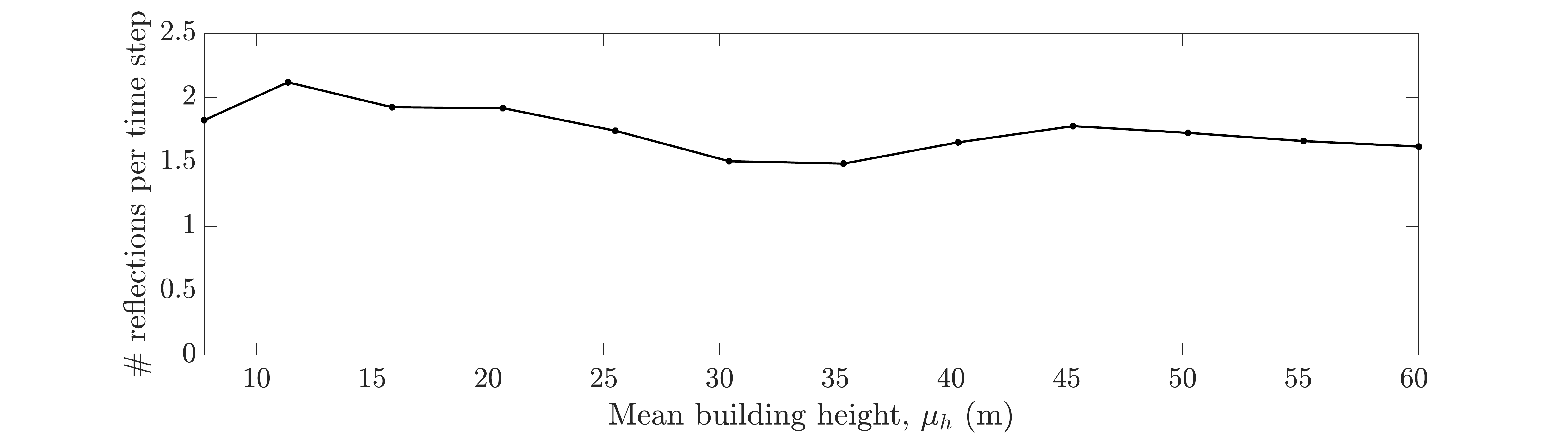}
	\caption{In general, fewer reflections are received in deeper urban canyons.}
	\label{fig:noRefls}
\end{figure}

\begin{figure}[t]
	\centering 
	\includegraphics[trim=130 0 125 0,clip,width=0.89\textwidth]{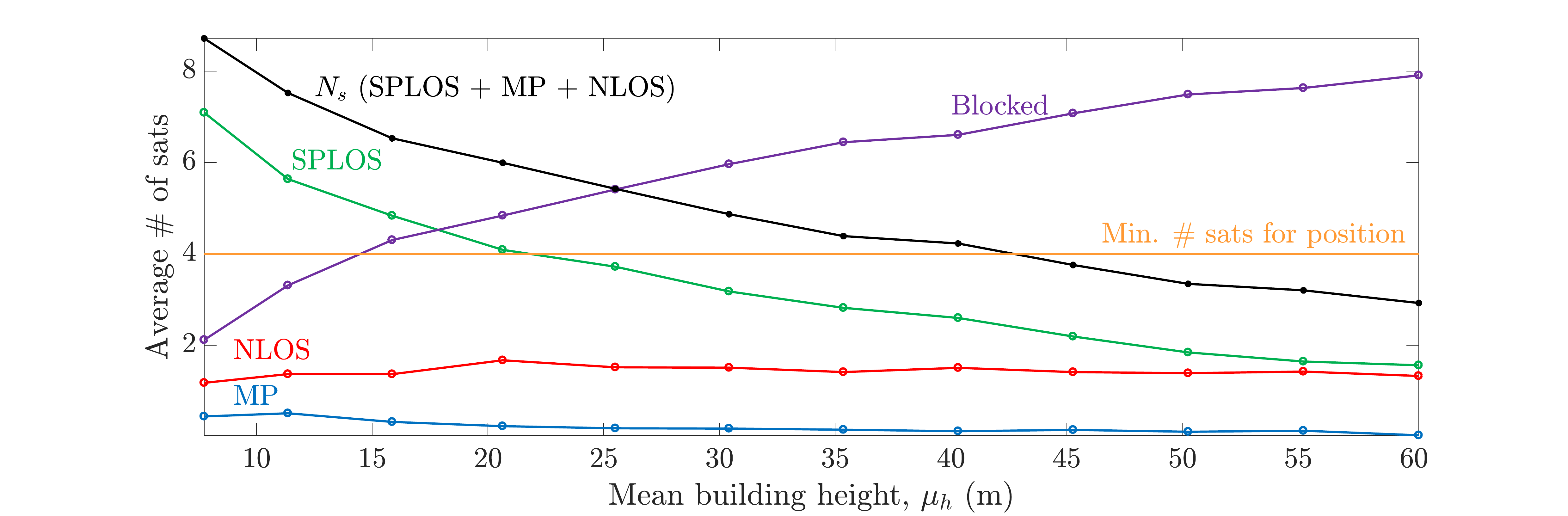}
	\caption{For deeper urban canyons, more satellites are blocked instead of being received as single path line-of-sight.}
	\label{fig:recModes}
\end{figure}

First, it can be noted that the total number of reflections per time step tend to decrease with increasing mean building height, see Fig.~\ref{fig:noRefls}. 
Second, with increasing values of building height, the satellites tend to switch from the category SPLOS to blocked, see Fig.~\ref{fig:recModes}. The multipath reception mode slightly decreases in frequency with urban canyon depth, while the NLOS mode becomes slightly more common. The reason is that increasing building height means not only that more satellites are blocked, but also that more reflections are blocked. If the simulation would allow the signals to reflect more than once these results could change, however these secondary reflections could be too weak to influence the receiver. The total number of received satellites $N_s$ is the sum of the modes SPLOS, MP, and NLOS. Somewhere between $\mu_h = 40 \, \text{m}$ and $\mu_h = 45 \, \text{m}$, less than four satellites are received, meaning that stand-alone 3D GNSS positioning becomes impossible.

\subsection{Reflection delay distributions}
\begin{figure}[t]
	\centering 
	\includegraphics[trim=120 0 140 0,clip,width=\textwidth]{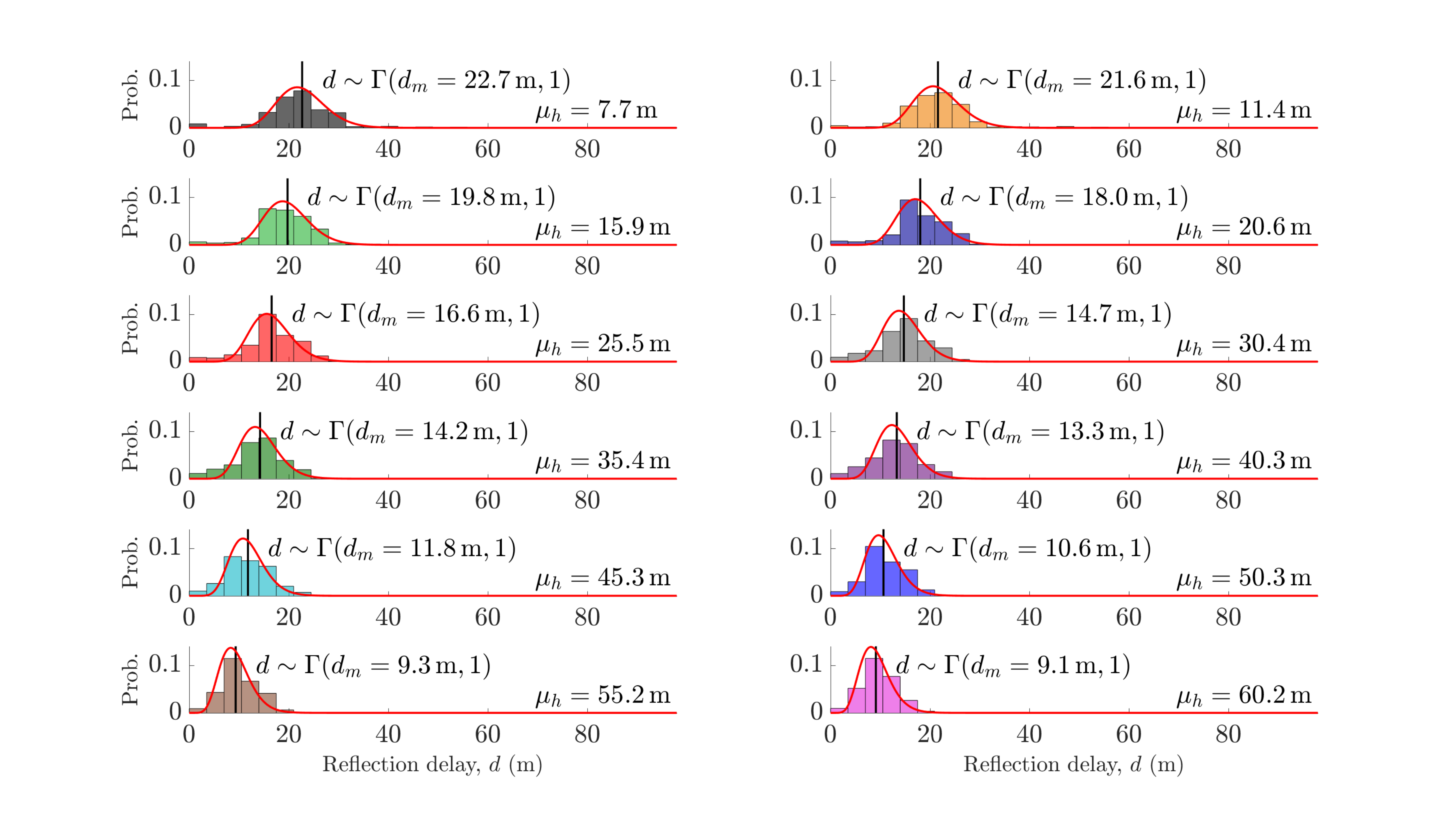} 
	\caption{In an open environment, the reflection delays are typically greater than in the deeper urban canyons. A few outlier reflections reach delay values of up to $350 \, \text{m}$, but they are not included in the figure, to make the delays below 100~m more visible.}
	\label{fig:delayPDF}
\end{figure}

As found by~\cite{Wang18}, a gamma distribution seems to fit the reflection delays well, see Fig.~\ref{fig:delayPDF}. In difference to~\cite{Wang18}, where the scale parameter is between 2 and 3, the scale parameter is 1 in our results. One reason could be differences in geometry, such as the wider streets in Shanghai or a larger value of $\sigma_h$. Another possibility is that this simulation only allows signals to reflect once, whereas~\cite{Wang18} interpreted the measured signals to have been reflected between 2 and 3 times on average. A further difference to~\cite{Wang18} is that the simulated delays are in general shorter, which could be attributed to the fact that the simulation is a symmetric urban canyon, while the Shanghai streets are not always blocked by buildings on both sides.

When the scale parameter of the gamma distribution is equal to 1, the shape parameter $d_m$ is approximately equal to the median reflection delay, which decreases with increasing mean building height. A few outliers delays exceed $300 \, \text{m}$, which is why the median is used instead of the mean. For deeper urban canyons with higher buildings, only high-elevation satellites can be reflected, and these satellites create smaller reflection delays. As seen in Fig.~\ref{fig:noRefls}, the higher values of $\nu_h$ result in fewer reflections, meaning that the empirical probability densities for these environments are less representative. 

\subsection{Model to estimate reflection delay distributions}
\begin{figure}[t]
	\centering
	\includegraphics[width=0.9\textwidth]{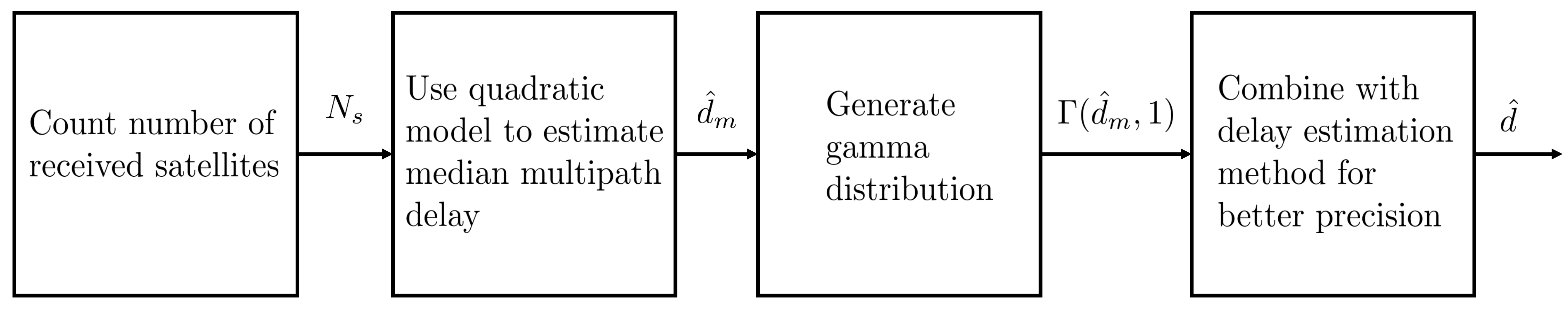}
	\caption{By counting the number of satellites, passing this parameter in the quadratic model, and the using the result as the shape parameter of a gamma distribution, the reflection delay distributions are estimated.}
	\label{fig:flowchart}
\end{figure}

\begin{figure}[t]
	\centering 
	\includegraphics[trim=120 0 130 0,clip,width=0.9\textwidth]{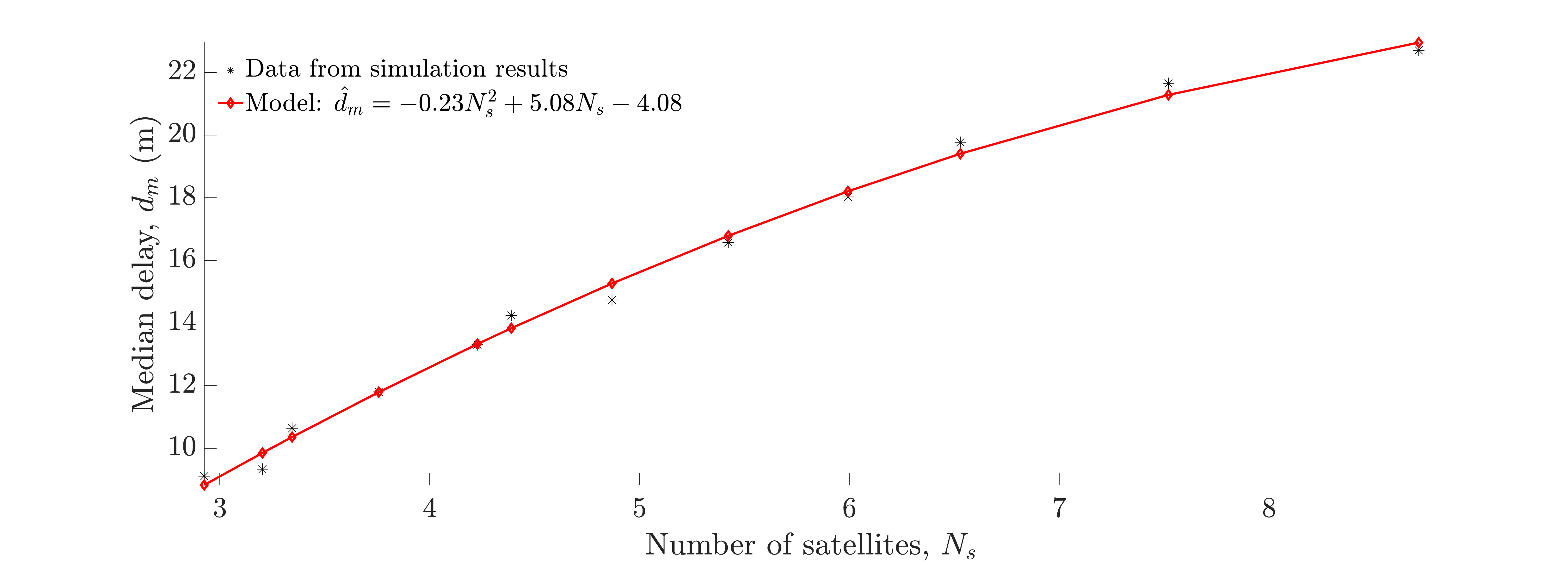}
	\caption{The median reflection delay can be estimated using a quadratic function of the number of received satellites.}
	\label{fig:model}
\end{figure}

The mean building height could be estimated using, e.g. 3D maps of the environment, however the number of satellites $N_s$ is always known by the receiver. We have that $N_s$ decreases with the mean building height. Likewise, the median reflection delay decreases with increasing building height. Thus, we can use the number of received satellites to estimate the median reflection delay, see Fig.~\ref{fig:flowchart}
\begin{align}
\hat{d}_m = -0.23 N_s^2 +5.08 N_s -4.08 .
\label{eq:model}
\end{align} 
This quadratic model was fitted using the least-squares algorithm, see Fig.~\ref{fig:model}. Its RMS estimation error of the median delay is
\begin{align}
\sqrt{\frac{1}{N}\sum_{n=1}^{N}(\hat{d}_{m,n} - d_{m,n})^2  } = 0.33 \, \text{m}.
\label{eq:modelerr}
\end{align} 
This applies to GPS only receivers, for multi-constellations receivers with, e.g., GPS and Galileo, the coefficients of~\ref{eq:model} might need adjustment.

\section{Conclusion}
As a conclusion, deeper urban canyons result in fewer reflections and shorter multipath delays. The delays follow a gamma distribution, whose shape parameter represents the median reflection delay. A simple quadratic model using the number of received satellites can estimate this median reflection delay. For the future, this information should be used in combination with multipath delay estimation methods, for improved performance. Furthermore, diffraction should be considered. Finally, it would be interesting to derive the median reflection delay based on the satellite elevation, which is a parameter that is easy to access. This could lead to more general results, that also apply to, i.e., unsymmetrical urban areas.

\appendix

\section{Appendix: Simulation Source Code}
The source code of this simulation is available at \url{https://github.com/Fusion-Goettingen}, for use as a benchmarking tool of multipath-resistant positioning algorithms. All the parameters described in this paper can be freely defined by the user. The satellite orbits were generated using "Satellite Constellation"~\cite{Riya18} (modified to support a custom sampling time and a custom start time of the day to allow fast-forwarding of the satellite constellation).

\vsp{1em}
\bibliographystyle{IEEEbib} 
\bibliography{IEEEabrv,MyListOfPapersENC}

\end{document}